# New class of compounds – variators – are reprogramming substrate specificity of H3K4me3 epigenetic marks reading PHD domain of BPTF protein.

Oleksandr Ya Yakovenko[*,¶], Sreeja Leelakumari[¶], Ganna Vashchenko[⌐], Pierre Cheung[¶], Albert Badiong[¶] and Steven J.M. Jones[¶]

[¶]*Canada's Michael Smith Genome Sciences Centre, BC Cancer Agency, Suite 100 570 West 7th Avenue, Vancouver, British Columbia V5Z 4S6, Canada*

[⌐]*Department of Biochemistry and Molecular Biology, University of British Columbia, 2350 Health Sciences Mall, Canada*

[*] *Corresponding author, email ayakovenko@bcgsc.ca*

*Abstract.*

In lymphoma, mutations in genes of histone modifying proteins are frequently observed. Notably, somatic mutations in the activatory histone modification writing protein MLL2 and the repressive modification writer EZH2 are the most frequent. Gain of function mutations are typically detected in EZH2 whilst MLL2 mutations are usually observed as conferring a homozygous loss of function. The gain-of-function mutations in EZH2 provide an obvious target for the development of inhibitors with therapeutic potential. To counter the loss of functional MLL2 protein, we computationally predicted compounds that are able to modulate the reader of the corresponding modifications, BPTF, to recognize other forms of the histone H3 lysine 4, instead of the tri-methylated form normally produced by MLL2. By forming a synthetic triple-complex of a compound, the histone H3 tail and BPTF we potentially circumvent the requirement for functional MLL2 methyl-transferase through the modulation of BPTF activity. Here we show a proof-of-principle that special compounds, named variators, can reprogram selectivity of protein binding and thus create artificial regulatory pathways which can have a potential therapeutic role. A therapeutic role of BPTF variators may extend to other diseases that involve loss of MLL2 function, such as Kabuki syndrome or the aberrant functioning of H3K4 modification as observed in Huntington disease and in memory formation.

*Introduction.*

Follicular lymphoma (FL) and diffuse large B-cell lymphoma (DLBCL) are two common non-Hodgkin lymphomas [1]. These cancers recurrently harbor somatic mutations in genes important for epigenetic regulation including *MLL2*, *EZH2*, *CREBBP* and *EP300* [2, 3]. The most frequently mutated (89% and 32% of FL and DLBCL cases respectively) is the MLL2 gene, which encodes a histone methyl-transferase. The known cellular function of MLL2, which is performed by its only catalytic domain SET [4] (amino-acid residues 5396-5517), is methylation [5] of the 4-th lysine of histone H3 (H3K4) — an epigenetic modification that is strongly correlated with actively expressed genes. This modification does not induce transcriptional activation on its own, but provides H3K4 marked chromosomal regions for binding of the selective marks reader — BPTF (the mark is recognized by its PHD finger domain) [6-8]. The affinity of BPTF to the different methylation states of H3K4 is significantly shifted toward the highly-methylated ones. According to data obtained using surface plasmon resonance [8] and subsequently confirmed with fluorescence polarization methods [9] the affinity is $K_d \approx 230$ μM for K4me0, $K_d \approx 46$ μM for K4me1, $K_d \approx 6$ μM for K4me2 and $K_d \approx 3$ μM for K4me3. This selective affinity of BPTF toward highly-methylated states is opposed to L3MBTL1 protein, which has higher affinity to the low-methylation states [10, 11] ($K_d > 500$ μM for K4me0, $K_d \approx 6$ μM for K4me1, $K_d \approx 76$ μM for K4me2 and $K_d > 500$ μM for K4me3) [12], and in combination allows cells to efficiently distinguish the activatory H3K4me3 modification and perform region specific chromatin remodeling.

The observed *MLL2* mutations in tumors indicate a strong preference for insertions, deletions and truncations consistent with homozygous loss of its catalytic activity in the cells. The result is an epigenetic transcriptional repression that is assumed to be pro-oncogenic possibly due to inhibiting of terminal differentiation pathways and maintaining the cells in a (semi-)proliferative state. In the absence of fully functional MLL2 protein, the H3K4me1 state is likely present in excess being produced by the SET7/9 methyltransferase (*de novo*) [13, 14] and from the pool of already existing highly-methylated states of H3K4 by demethylases KDM1B (from H3K4me2) [15] and JARID1C/SMCX (from H3K4me3) [16] while amount of highly-methylated H3K4 states is decreased due to the loss of MLL2 activity. The reduction in highly methylated H3K4 states diminishes the ability for BPTF to appropriately recognize and bind to the chromatin rendering it as a generally ineffective transcriptional activator which function is even further suppressed by competition with L3MBTL1. At the same time repressive histone marks continue to be laid down by the polycomb-group of proteins (pcG) and this rate is often increased as its catalytic component, EZH2, [17, 18] (the writer of H3K27 methylated marks) is often further activated through gain-of-function mutations [19, 20]; EZH2 have been determined as the second most frequent events affecting epigenetic regulation in lymphomas [2, 3]. For EZH2 there has already been significant effort in the development of inhibitors as potential therapeutics [21, 22]. Conversely, in this study we decided to alter the function of the H3K4me3 reader, BPTF, as a way to counteract the loss of functional MLL2 protein. Our goal was to reprogram the binding affinity of BPTF to H3K4me1 through the formation of a triple-complex of H3K4me1, BPTF and a compound that would strengthen this interaction compensating for the lack of functional MLL2.

The structure and mechanism of action of the methylated lysine binding site of PHD domain of BPTF protein has been previously determined [9, 12]. It is a classic aromatic cage able to sequester the triple methylated ammonium cation of methylated lysine side-chain through cation-π interactions [9, 12, 23]. The cage is tuned to strongly bind hydrophobic triple- and double- methylated aliphatic amines — it has a cubic shape and four out of its six walls are formed by aromatic residues Y10, Y17, Y23 and W32 (of the two remaining facets one is opened to water and the last one is where the Lysine hydrocarbon side-chain comes from). Li et al. [12] had previously modified the aromatic cage of BPTF with site-directed mutagenesis to reinforce the H3K4me1 interaction from Kd=46 μM in the wild-type to $K_d$=13.1 μM for the Y17E mutant (Y17D and Y17Q were less efficient reprogrammers). In this work we have reprogrammed the affinity of BPTF to the lower methylation states of H3K4 mark with small organic compounds. The identified compounds show selective cytotoxicity against *MLL2* mutant cancer cell lines and modulate binding of BPTF to its substrates in pulldown experimentsl.

*Methods.*

*Strategy of rational design.* A two-stage computational protocol was developed to identify promising candidates for the reprogramming of BPTF affinity (the flowchart of our virtual screening protocol is shown on Fig 1). At the first stage virtual compounds – referred as probes - that stabilize all desired interaction in molecular mechanical model were developed. At the second stage real substances that mimic the desired behavior were identified using knowledge about favorite protein conformations that were induced by the probes at the first stage. Despite the probes were used only for preparation purposes, so that their physical existence is not required, all probes were designed as chemically feasible substances to obtain even more realistic models of bptf reprogramming.

*Triple complex preparation.* Generally speaking there are three *a priori* important features of any reprogramming compound. Firstly it should contain an 'anchoring fragment' that is responsible for binding of the compound to BPTF. Because there are no empty pockets in proximity of H3K4 binding site of BPTF, we decided to disassemble one wall of the aromatic cage and use the freed room as the anchoring site for a candidate compound. The individual importance of each of the four cage forming residues has been previously measured using fluorescence polarization methodology by determination of dissociation constants of H3K4me3 peptide to different BPTF mutant proteins [9]: $K_d$WT=1.6 μM, $K_d$Y10T=40.3 μM, $K_d$Y17T=10.0 μM, $K_d$Y23S=59.7 μM and $K_d$W32F=34.4 μM. These data indicate that the least evolutionary conserved Y17 residue has the smallest contribution to substrate binding and thus it was chosen for disassembly and substitution with the anchoring fragment of a probe. The anchoring fragment thus would likely be an aromatic or aliphatic ring(s) as it occupies the cavity of Y17 side-chain and is a facet of the aromatic cage. Secondly, in order to manifest reprogramming effect, the compound has to interact with the mono-methylated lysine to stabilize it in the active site of BPTF. It is thus assumed to be a 'reprogramming motif' possessing a negatively charged hydrogen bond accepting pattern to strengthen the binding to low methylated forms of H3K4 in the hydrophobic environment. In our case a ketone group is expected to be superior than previously reported carboxyl reprogramming motif of the Y17E mutant [12] because it has smaller solvation free energy and therefore less likely to disturb the affinity of the anchoring fragment to BPTF. The third desirable feature is that the compound should not compete with the binding of low-methylated lysine tail to BPTF but stabilize the complex.

The feature was estimated using molecular dynamic (MD) simulations of a triple complex of BPTF, H3 histone tail and a probe. The triple complexes of a probe bound to BPTF with H3K4me1 peptide were derived from the model of the binary complex resolved by Li *et al.* using NMR methodology [9] (www.pdb.org, accession code 2FUU). Conformation of the Y17 side-chain was manually edited through rotations around dihedral angles to cause the hydroxyl-phenyl ring to leave the cavity occupied in the original model and was located at the opposite side of the site relative to the Y17 CA atom. Although the free energy of rotation around the dihedral angle in binary complex of BPTF with H3K4 tail can be rigorously evaluated with MD it is not relevant for the triple complex because Y17 side-chain is assumed to interact strongly with the reprogramming compound which structure is not yet know. Due to this interaction, to the three general *a priori* known features of the reprogramming substance we added the forth one (specific) – a candidate has to bear a motif that stabilizes 'open' conformation of Y17 residue. The first minimal probe was simply composed accordingly to four *a priori* conditions described above as a virtual compound which consists of a bare aromatic ring attached to a reprogramming ketone group and an ethyl group to retain Y17 outside of its natural pocket. The probe was manually inserted into the empty cavity of Y17 sidechain by its anchoring fragment and the stability of the triple complex was evaluated with MD and then reviewed. Based on our observations of the loss of stability within the MD simulation, the structure of the probe was mutated to evolve into a more stable triple-complex and the iteration was repeated. After 6 iterations the probe (V6) that conferring stable behavior of the triple-complex with BPTF and the H3K4me1 tail peptide in the MD simulations was determined.

*Detailed MD protocol*. Prior to the simulations every model was placed in triclinic box of simple point charge (SPC) water [24] into which 100 mM NaCl equivalent was added including neutralizing counter-ions. Periodic boundaries were applied in all directions. The N- and C-termini of all protein molecules were ionized. All other amino acids were assigned their canonical state at physiological pH. Energy terms from the GROMOS96 43a1 parameters set[25] were applied to all molecular species in the system. To handle two $Zn^{2+}$ ions the parameter set was updated with eight new coordinatechemical bonds (and corresponding angles): for the first zinc atom to SG atom of C11, SG atom of C13, ND1 atom of H34 and SG atom of C37; for the second zinc atom to SG atom of C26, SG atom of C29, SG atom of C53 and SG atom of C56. The charge distributions on corresponding residues were updated either: for the first $Zn^{2+}$ coordination complex one electron was smoothed along the H34 imidazolering and the second was divided equally between the zinc atom and three sulfur atoms of cysteine residues; for the second zinc atom charge was set as -0.4e and the remaining two electrons were equally divided among four sulfur atoms of the cysteine residues. The K4me3 residue of the H3 histone peptide was edited to replace two of the three methyl groups with hydrogen atoms. Due to absence of accurate parameters for cation-π interactions in GROMOS96 force field all atoms within aromatic rings of the BPTF aromatic cage, but not those of the Y17 residue, as well as the NZ atom of the mono-methylated lysine 4 were restrained at their coordinates in the 2FUU model with 1000 kJ·mol$^{-1}$·nm$^{-2}$ as hard harmonic position restraints. All triple complexes were constructed manually by rotation, translation and dihedral angle alternations of compounds which were optimized in vacuum at PRODRG server [26]. Due to some lack of accuracy in parameters sets generated by PRODRG [27] the non-bonded parameters of compounds were intensively edited manually, especially the charge distribution which was rewritten from scratch for each compound.

All MD simulations were carried out with somewhat 'stickier' non-bonded interactions: both Leonard-Jones and Coulomb short range interactions were switched at 1.3nm and vanished at 1.4nm with neighbor-searching radii set to 1.5 nm and repeated for every 10 steps of the MD integrator. Long-range electrostatic interactions were modeled with the particle mesh Ewald (PME) algorithm [28]. To model solvated complexes, the structures were relaxed by l-BFGS minimization [29] and 50picoseconds(ps) of molecular dynamic simulations with restrained positions of proteins' and compounds' heavy atoms were simulated under a constant volume (NVT) ensemble. A simulating annealing [30, 31] was used to warm up the system from initial velocities assigned accordingly to Boltzmann distribution at T=10K till T=310K at the end. Following NVT warm up, 100ps of constant pressure (NPT) equilibration was performed. Complex and solvent with ions were coupled to a separate temperature coupling baths and the temperature was maintained at T=310K. For equilibration a weak coupling method[32] was used to maintain pressure isotropicallyat 1.0 bar and temperature constant at 310K. All subsequent productive runs (of 12 ns each) were performed with the more accurate Nose-Hoover thermostat [33, 34] with temperature coupling time constant of 0.1ps and the Parrinello-Rahman barostat [35, 36] with pressure coupling time constant of 1.0ps under NPT ensemble. This combination of thermostat and barostat ensures that a true NPT ensemble is sampled and the fact of restraining of a tiny (22 atoms) and compact aromatic cage causes only small artifacts if any. For visual inspections of MD trajectories VMD viewer [37] was used.

*Virtual screening.* To identify real compounds that could replace our evolved probe we took the average protein structure from the stable MD with probe V6 and used it for docking of the entire NCI plated database (http://dtp.nci.nih.gov). High-throughput virtual screening was carried out with molecular docking method implemented as a part of ICM package [38, 39] with its default parameters. C.a. 5K highest-scored triple complexes were reviewed manually and the only docked compounds that were similar to that of stable probe V6 and met all *a priori* known reprogramming requirements were considered as promising. Each time a promising compound was encountered during the visual inspection of the docking results, the inspection was interrupted; the triple complex of new candidate was created and evaluated with MD as it is described above for virtual probes. Meanwhile the MD was evaluated, even more promising structures similar to the encountered one were searched through entire database using substructures search implemented at https://pubchem.ncbi.nlm.nih.gov/. All interesting homologues were evaluated manually, regardless of the molecular docking scores they possessed, to consider all highly similar molecules that were not scored well by the molecular docking software. The stability of the triple complex was instead speculated from the behaviour of the previously simulated complexes (the amount had been increasing through the selection process). In all promising cases and the cases where it was difficult to make the conclusion due to diversity of the new candidate from all those that had been already simulated, the stability of its triple complex was either evaluated with the same MD protocol. MD trajectories were then reviewed and conclusions about the importance of particular substitutions were made to improve the consequent continuation of the visual inspection of docking results. Finally 22 compounds that provided a stable triple-complex behavior in MD simulations were chosen for testing *in vitro* with lymphoma-derived cell lines.

*Cell Culture and Proliferation Assays.* The DLBCL cell lines were maintained in RPMI medium 1640 (Life Technologies) supplemented with 10% (v/v) fetal bovine serum (Life Technologies) and 1% penicillin/streptomycin (Life Technologies). All cell lines were grown in a 37°C incubator with 5% $CO_2$, and humidified atmosphere. Powdered drug compounds were solubilized and maintained in the drug-carrier, dimethyl sulfoxide (DMSO), at a concentration of 10mM. In preparation for cell treatment, these drug stocks were diluted in RPMI 1640 medium. Cells to be treated were harvested and made up to a suspension with a density of $4\times10^5$ cells/mL. 90μL of this suspension were placed into each well of a MICROTEST™ 96-well Assay Plate, Optilux™ (BD). 10μL of the drug solution was added to each well as the treatment. Also included in each plate were a background noise control (medium only), untreated cells control, and a drug-carrier (DMSO) control. Cells are kept in a 37°C incubator with 5% $CO_2$, and a humidified atmosphere with their treatment conditions for 48 hours. Proliferation of drug-treated cells was measured by incubating them with 10% alamarBlue® (Life Technologies) for 2 hours. Fluorescent signals were later measured, the background noise subtracted, and then normalized to the drug-carrier control's signal.

*Protein Preparation.* The full-length human BPTF construct was obtained from C. David Allis of the Laboratory of Chromatin Biology, Rockefeller University [9]. Primers were designed to capture the BPTF bromodomain and PHD domain. The PHD domain is responsible for recognizing H3K4me3 marks while the bromodomain is responsible for recognizing histone H4K16ac. The dual PHD finger-bromodomain (residues 2583-2751) from the human BPTF (gi:31322942) was cloned into a pDEST15 vector using Gateway® cloning technology allowing for an N-terminal GST tag (Life Technologies). Over-expression of the GST-BPTF dual PHD finger-bromodomain was induced in BL21-AI™ chemically competent *E. coli* cells (Life Technologies) using LB medium supplemented with 1 mM IPTG (Santa Cruz Biotechnologies) and 0.2% L-arabinose (Sigma) for 2 hours in a 37°C shaking incubator. GST-BPTF dual PHD finger-bromodomain was purified using Glutathione Sepharose 4B media (GE Healthcare Life Sciences) and dialyzed against PBS (Life Technologies) overnight.

*PullDown Assay.* The peptide pulldown experiment was performed essentially as previously described [40]. Briefly, biotinylated H3K4 peptides with methylation states ranging from me0 (no methyl groups) to me3 (3 methyl groups) were purchased from AnaSpec. These peptides were immobilized on streptavidin dynabeads (Dynabeads® M-280 Streptavidin by Invitrogen™) and were used to bait the GST-tagged BPTF into binding. Fifty microlitres of M-280 streptavidin-coupled Dynabeads® (Life Technologies) were dispensed into 1.5 mL microtubes and washed 3 x 50 µL in PBS. The beads were then incubated with C-terminally biotinylated H3K4me0, me1, me2 or me3 peptides, 21 amino acids long, rotating for 1 hour at 4°C under saturating conditions. The beads were washed 3 x 100 µL in HBS-TD (10 mM Na-HEPES, 150 mM NaCl, 0.005% Tween-20, 2 mM DTT) and incubated with 8.2 µM of GST-BPTF along with either 82 µM of drug in a 1:100 DMSO to PBS solution or carrier alone rotating for 3 hours at 4°C. The beads were then washed 10 x 200 µL in HBS-TD. The protein was eluted using 2x LDS sample buffer (Life Technologies) and 1x reducing agent (Life Technologies) at 85ºC for 10 minutes. The eluted samples were analyzed by western blot using a GST antibody (Santa Crus Biotechnologies) in order to visualize the GST-BPTF protein. Quantification of bands was made using the Image J software.

*Results.*

*Cell-based activity assay identified highly potent chemical structures.*

As a reporter for *in vitro* screening of BPTF reprogramming compounds, *MLL2* homozygous insertion/deletion mutants SU-DHL-9 [41] cell lines were used and the control was *MLL2* wild-type cell line, DoHH2 [42]; both are diffuse large B-cell lymphomas (DLBCL). The screening of the 22 computationally selected compounds revealed two promising scaffolds (represented by NSC382001 and NSC127763 compounds) with differential cytotoxicity to *MLL2* mutant cells versus *MLL2* wild-type cells. Due to abundance of already existing derivatives, one of the promising compounds (NSC382001) was chosen as the "leading" compound for future optimization. Approximately 70 more compounds, which were structurally similar to NSC382001, from the NCI plated and Otava Ltd (Otava Ltd., Toronto, ON, Canada, http://www.otavachemicals.com/) collections were subjected to a new round of cell-based testing. In this case "promising" compounds were selected mainly using structure-activity relationship (SAR) and only limited MD modeling. This optimization process provided two diverse but potent (Fig2.b.) inhibitors of tumor proliferation NSC304107 and 5980483 (Fig2.a). The four compounds had successfully passed PAINS filters for probable false-positive hits [43] as it is implemented at http://cbligand.org/PAINS/search_struct.php.

*Assessment of compounds in their ability to reprogram the affinity of BPTF to H3K4.*

The formation of the triple complex and the ability of the compounds to alter the affinity of BPTF toward differentially methylated H3K4 tail peptides were confirmed with a set of pulldown experiments.

The pulldown assay confirms the reprogramming effect of the compounds altering the stability of BPTF complexes with peptides in different methylated states of H3K4 residue (Fig 3.). It is observed that the presence of NSC304107 slightly increases the interaction of BPTF with H3K4me1 but decrease the interaction with H3K4me3 while 5980483 significantly stabilizes BPTF binding to H3K4me0, H3K4me1 (especially) and even to H3K4me2. The pulldowns ultimately indicates that both variators change the physical and chemical microenvironment of the active site of BPTF presumably with the hydrophilic h-bond accepting ketone group which makes the site less hydrophobic and more suited for substrates with (two) h-bond donors. 5980483 illustrates the changes via significantly increased binding to peptides with methylation states that harbor at least one polar hydrogen on their K4 residue. NSC304107 is different — the perturbation it applies makes the site of BPTF less suited for the natural substrate — H3K4me3, which has no polar hydrogens. Consequently, the hydrophobic H3K4me3 is incompatible with the triple complex formation with NSC304107 and competes with it for BPTF. Though the modality of the two reprogrammers is different, the pulldown results provide a strong evidence in favor of the hypothesis of a triple complex formation in close proximity from BPTF active site as the mechanism of variators acting. Otherwise it is difficult to explain how 5980483 can selectively increase affinity for substrates that differs only by one or two methyl groups at the end of long K4 residue or how NSC304107 can decrease affinity to the strongest binder ($K_d \approx 3\mu M$ for H3K4me3) whilst it increases the affinity to the weaker ones ($K_d \approx 46 \mu M$ for K4me1). It is notable that the similar (to 5980483) outcome was observed in the case of the second of identified promising scaffold (NSC127763) for which we were not able to optimize due to absence of sufficient amount of existing structural homologues but for which the mechanism of action was also successfully confirmed through the pulldown assay.

*Discussion.*

Here we provide a proof-of-principle that we can synthetically reprogram the affinity of proteins to an unnatural substrate. In particular, an artificial functional linkage between BPTF and the lower methylation states of H3K4 was created. The reprogramming compounds, named variators, modify the normal activity of the protein and provide novel drug design opportunities that are of special interest for compensating for the loss-of-function mutations within tumor suppressor genes, as we demonstrate for MLL2 deficiency. The treatment of lymphoma targeting a transcriptional activation program, normally determined by functional MLL2, is attractive as this appears to be a process abrogated at a higher frequency than the hyper-activity of repressive mark writing by EZH2 [2, 3, 44]. This increased frequency is either due to the fact that it is stochastically more likely for lymphoma to accrue an inactivating mutation in *MLL2* than a specific gain-of-function in *EZH2* or due to better tolerance of perturbing the activatory versus repressive branches of epigenetic regulation for cellular survival.

The general paradigm in drug development typically has been to identify inhibitory compounds and it is likely inherently obvious that it is easier to develop an inhibitor than a compound that would modulate the function of a protein. Therefore, we believe that our approach to develop a modulator of BPTF is of interest and provides an example of state of art reprogramming which affects only specific pathway where a compound can compensate for a loss-of-function mutation elsewhere in a pathway. We also described a computational approach for the identification of initial compound candidates that specifically attempts to identify up-front the modulating behavior that may be otherwise difficult to achieve using high-throughput compound screening approaches.

Though the perturbing of the normal protein functions is *a priori* toxic for a cell, the toxicity can be gained in such a way so it is more expressed against targeted tumor genomes. Invoking an artificial or synthetic pathway linkage between BPTF and H3K27me0 and H3K27me1 is suggested to preferentially target tumor over normal cells — in the presence of functional MLL2 in non-cancer cells, the abundance of low methylated H3K4 histone tails is significantly decreased. Another potential advantage of using of a reprogramming compound is to instigate a transcriptional program that can harbor non-synonymous somatic mutations and are accrued within these now activated genes. This may provide antigenic targets otherwise hidden due to epigenetic regulation [45], that could elicit an immune response against the cancer cells.

A role for such BPTF variators may extend to other diseases beyond cancer. Kabuki syndrome, a rare autosomal dominant development disorder, that is caused by the heterozygous loss of MLL2 presumably through the diminished activity of MLL2 [46]. Recently, a reduction of H3K4me3 has been observed in Huntington disease effected cells the mitigation of which reversed the disease associated down-regulation of neuronal genes [47]. MLL2 has also been found be required for long-term memory formation in mice [48] therefore it is possible that these compounds may ameliorate symptoms for some degenerative brain diseases.

*Conclusions.*

The computational rational design of conceptually new class of drugs — variators — is illustrated with BPTF protein reprogramming compounds. Variators, as opposed to widely developed inhibitors, do not inhibit existing biochemical processes but create new pathways *in vivo* by instigating novel molecular interactions. With this approach we were able to provide evidence that we can compensate for loss-of-function mutations in MLL2 and the absence of their H3K4me3 activating epigenetic marks which is frequently observed in lymphoma.

*Acknowledgements.*

Authors thank Dr Patel of Memorial Sloan-Kettering Cancer Center, New York, and Dr. C. David Allis of the Laboratory of Chromatin Biology, Rockefeller University for the full-length human BPTF construct.

Authors are grateful to the Developmental Therapeutics Program, National Cancer Institute, USA, for providing the chemical compounds in this project.

*Figures and Legends*

**Fig 1. The virtual screening protocol used for identification of BPTF modulators.**

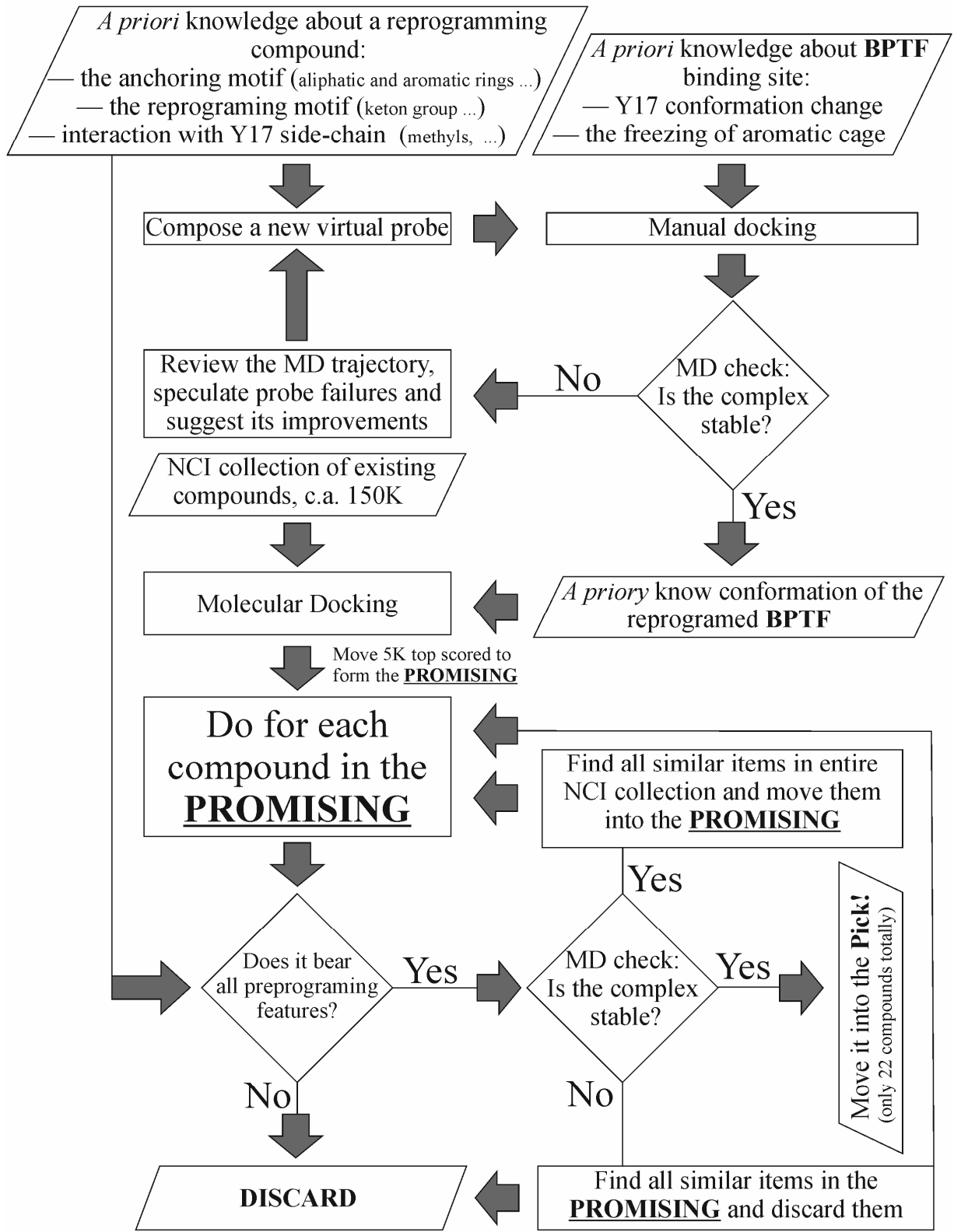

Our virtual screening protocol essentially consists of two loops: a) probe design loop, where the virtual modulator is developed and confirmed to work in the model, and b) modulator identification loop, where a set of real compounds is tested to mimic (in the model) the behaviour of previously developed modulator. The successful real candidates that worked as virtual modulators in the model were then tested experimentally. Please note however that our goal was identification of the modulators rather than methodology developing so the protocol was not strictly followed and sometimes violated. For example, in the case of a candidate that was able to preserve the triple complex through significant duration of MD simulation but dissociated later, its homologues from entire NCI collection were moved into the PROMISING even despite of the failure (yet the candidate itself was rejected).

**Fig 2. The model complex and cell line cytotoxicity of MLL2 reprogramming candidates.**

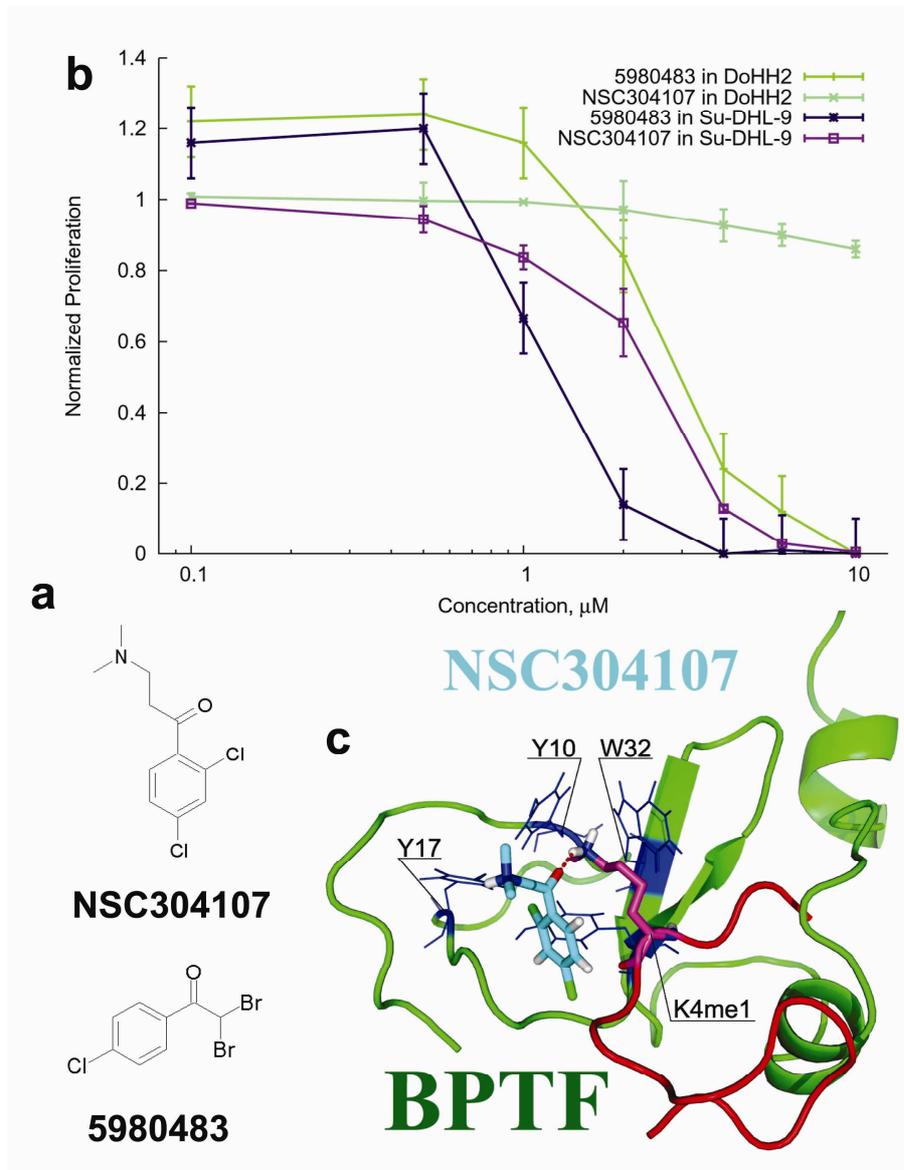

**a.** The chemical structure of the BPTF variators. **b.** Potency of the variators against *MLL2* homozygous deletion mutant (SU-DHL-9) and *MLL2* wild-type (DoHH2) cell lines as measured by Alamar Blue proliferation studies. **c.**

A 3D structural model of the triple complex including the variator (NSC304107), histone H3 peptide tail mono-methylated at K4 and BPTF protein. BPTF is shown in green with the residues of its aromatic cage as blue lines, the H3 tail is shown in red with the H3K4me1 residue in pink and NSC304107 is shown as cyan. The reprogramming hydrogen bond between H3K4me1 side-chain and ketone group of the variator is shown as a red dotted line.

**Fig3. Pulldown experiments confirm reprogramming activity of the active compounds.**

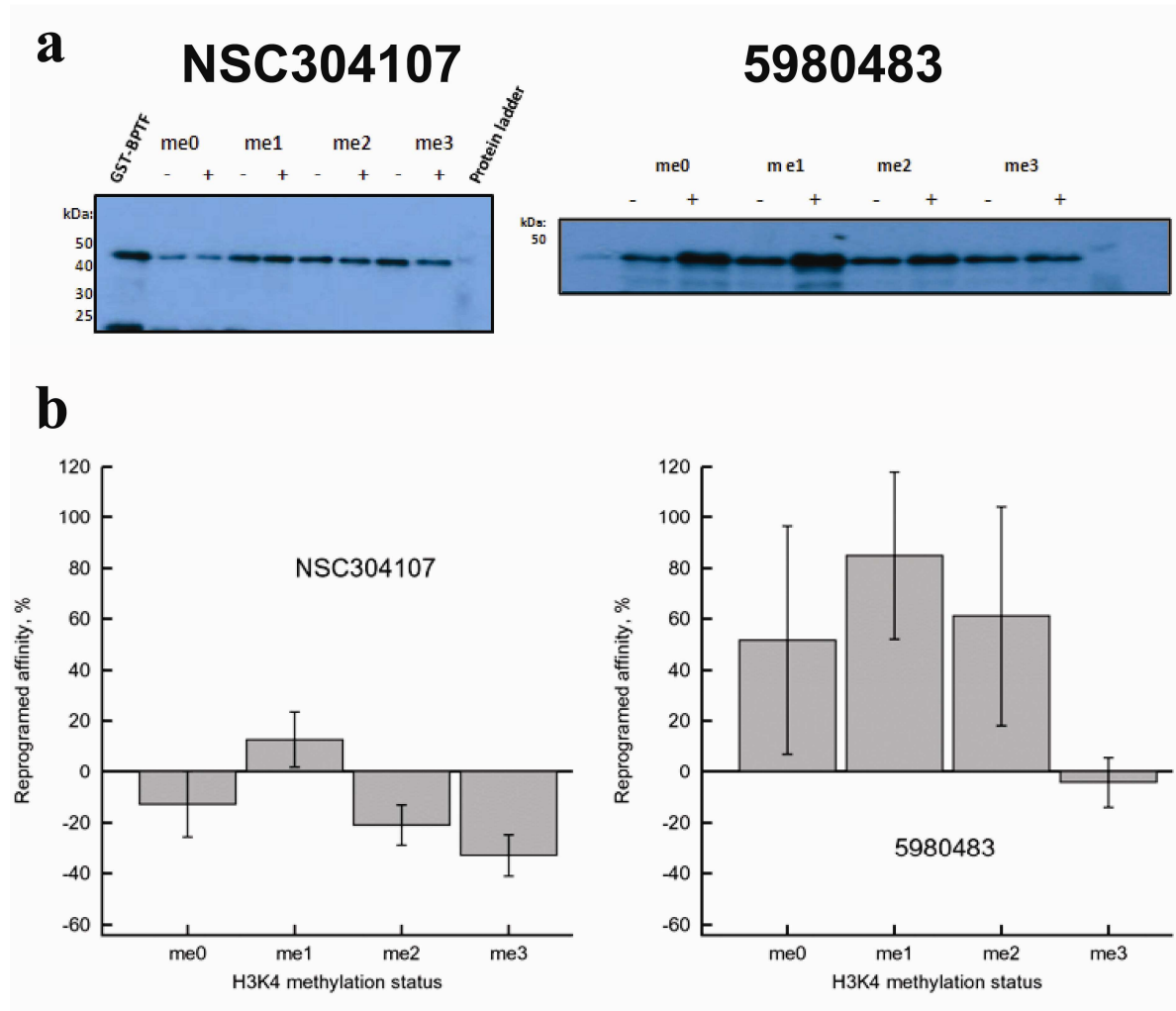

**a.** An example of actual pulldown gels; '+' and '-' indicate presence or absence of the compound. **b.** Averaged relative densitometry difference of stains from pulldown gels. Results are shown with mean and standard error from three independent experiments.